\documentclass[conference]{IEEEtran}
\IEEEoverridecommandlockouts
\usepackage{cite}
\usepackage{amsmath,amssymb,amsfonts}
\usepackage{algorithmic}
\usepackage{graphicx}
\usepackage{textcomp}
\usepackage{xcolor}
\usepackage{float}
\usepackage{tikz}
\usepackage[ruled,vlined]{algorithm2e}
\usepackage{hyperref}
\def\BibTeX{{\rm B\kern-.05em{\sc i\kern-.025em b}\kern-.08em
    T\kern-.1667em\lower.7ex\hbox{E}\kern-.125emX}}
    \usetikzlibrary{arrows}

\tikzstyle{int}=[draw, fill=blue!20, minimum size=2em]
\tikzstyle{init} = [pin edge={to-,thin,black},align=center]

\tikzset{
    pics/microph/.style={code={ 
        \draw[black, line width=.2em, rounded corners=1.7ex] 
            (-.85em,4.5ex) -- (-.85em,2ex) -- (.85em,2ex) -- (.85em,4.5ex);
        \fill[black] 
            (-.6em,5ex) to[rounded corners=1.2ex]  
            (-.6em,2.5ex) to[rounded corners=1.2ex] (.6em,2.5ex)
            -- (.6em,5ex) to[rounded corners=.2ex] ++(-.85em,0) to[rounded corners=.2ex] ++(0,.35ex) -- ++(.85em,0)  
            -- (.6em,5.5ex) to[rounded corners=.2ex] ++(-.85em,0) to[rounded corners=.2ex] ++(0,.35ex) -- ++(.85em,0)
            -- (.6em,6ex) to[rounded corners=.2ex] ++(-.85em,0) to[rounded corners=.2ex] ++(0,.35ex) -- ++(.85em,0)
            -- (.6em,6.5ex) to[rounded corners=.2ex] ++(-.85em,0) to[rounded corners=.2ex] ++(0,.35ex) -- ++(.85em,0)
            to[rounded corners=1.2ex]
            (.6em,8ex) to[rounded corners=1.2ex]
            (-.6em,8ex) to cycle; 
        \fill[black] (-.1em,1.8ex) rectangle (.1em,.5ex);
        \fill[black] (-.5em,.5ex) rectangle (.5em,0);
    }},
    Speaker/.pic={
        \filldraw[fill=gray!40,pic actions] 
        (-15pt,0) -- 
        coordinate[midway] (-front) 
        (15pt,0) -- 
        ++([shift={(-6pt,8pt)}]0pt,0pt) coordinate (aux1) -- 
        ++(-18pt,0) coordinate (aux2) 
        -- cycle 
        (aux1) -- ++(0,6pt) -- coordinate[midway] (-back) ++(-18pt,0) -- (aux2);
  }
}
\tikzstyle{int}=[draw, fill=blue!20, minimum size=2em]
\tikzstyle{init} = [pin edge={to-,thin,black},align=center]

\hypersetup{colorlinks=true,%
            breaklinks=true,%
            linkcolor=black,%
            pdfstartview=FitH,
            pdfpagelayout=TwoColumnRight}%
\begin{document}

\title{On the Integration of Acoustics and LiDAR: a Multi-Modal Approach to Acoustic Reflector Estimation\\
\author{Ellen~Riemens,
        Pablo~Mart\'inez-Nuevo,
        Jorge~Martinez,
        Martin~M\o{}ller,
        Richard~C.~Hendriks}}%
\maketitle

\begin{abstract}
% Loudspeakers are usually placed in an environment unknown to the loudspeaker designers. 
Having knowledge on the room acoustic properties, e.g., the location of acoustic reflectors, allows to better reproduce the sound field as intended.
Current state-of-the-art methods for room boundary detection using microphone measurements typically focus on a two-dimensional setting, causing a model mismatch when employed in real-life scenarios. Detection of arbitrary reflectors in three dimensions encounters practical limitations, e.g., the need for a spherical array and the increased computational complexity. Moreover, loudspeakers may not have an omnidirectional directivity pattern, as usually assumed in the literature, making the detection of acoustic reflectors in some directions more challenging. In the proposed method, a LiDAR sensor is added to a loudspeaker to improve wall detection accuracy and robustness. This is done in two ways. First, the model mismatch introduced by horizontal reflectors can be resolved by detecting reflectors with the LiDAR sensor to enable elimination of their detrimental influence from the 2D problem in pre-processing. Second, a LiDAR-based method is proposed to compensate for the challenging directions where the directive loudspeaker emits little energy. 
We show via simulations that this multi-modal approach, i.e., combining microphone and LiDAR sensors, improves the robustness and accuracy of wall detection.
\end{abstract}


\section{Introduction}
Room properties have a large influence on the sound field reproduction, meaning the highest reproduction quality can only be obtained when the room influence is known. For example, when a loudspeaker is placed close to a wall or corner, the lower frequency range in the room is amplified compared to the mid and high frequency, resulting in an unbalanced sound experience \cite{fuchs2015Requirement}. 
Working from home has led to a large increase in teleconference meetings. To use teleconference meetings as a feasible alternative to in-person meetings, speech intelligibility is a crucial factor. The reduced speech intelligibility due to echoes introduced by the room, makes it crucial to be aware of the nearby walls that introduce these echoes \cite{bradley1999Combined}. 
The introduction of smart loudspeakers gives rise to opportunities to estimate room parameters to improve the sound experience of the user. Of particular importance is information on the proximity of the walls. If the wall locations are known, their effects can be compensated for using digital filters \cite{spors2007Active}\cite{rabenstein2005Limiting}.
In this paper we address the problem of detecting the walls in close proximity to the speaker using sensors on the speaker.

Modern smart loudspeakers are often equipped with microphone arrays for voice commands. The presence of these microphone arrays can also be exploited for the detection of the nearby reflective surfaces together with the loudspeaker driver via the room impulse response (RIR). In \cite{dokmanic2013Acoustic}, Dokmani\'{c} et al. aimed to reconstruct a convex polyhedral room from impulse responses exploiting the properties of Euclidean distance matrices (EDMs) in the general 3D case. Since this combinatorial problem is NP-hard, the computational complexity is high. Furthermore, precise timing information on the impulses is required, which is challenging due to filtering. De Jager et al. proposed \cite{jager2016Room} a method that solves the combinatorial problem at the same accuracy but at a much lower complexity by posing the problem as a graph. Coutino et al. proposed a greedy method to further reduce this computational complexity in \cite{coutino2017Greedy}. Even though the recent algorithms are much faster, this is still a limitation. Furthermore, these methods rely on perfect peak detection from the RIR. More recently, Zacc\`{a} et al. \cite{zacca2021Inferring} posed this problem as a linear system. This system directly maps the image source locations to the received impulse response, including the loudspeaker directivity. This method does not rely on peak detection from the impulse response, however, the computational complexity of solving the inverse problem limits the method in practical applications to 2D. Furthermore, the successful identification of reflectors in regions where the loudspeaker radiates little energy, i.e., the back, is considerably lower than would be desirable.

It is possible to take advantage of the presence of other sensing modalities like LiDAR to detect walls more accurately. This could be done using point clouds that give direct depth information and can be used to detect planes.
Plane detection from point clouds is commonly approached through model fitting. The first category of popular model fitting approaches is Random Sample Consensus (RANSAC) \cite{fischler1981Random}\cite{schnabel2007Efficient}. A disadvantage of RANSAC is that when the number of iterations is limited, the solution may not be optimal. It relies on problem-specific parameters. In addition to that, using this approach the solution cannot be obtained analytically, which could be exploited when combining information of multiple modalities. In order to find multiple surfaces with RANSAC, the algorithm requires multiple runs and merging of detected planes. Considering these fundamental limitations of RANSAC approaches, this category is not further considered. The second popular category of model fitting approaches is based on the Hough Transform. This procedure is computationally very demanding, resulting in many adaptations of this algorithm.
The Probabilistic Hough Transform \cite{kiryati1991Probabilistic} and the Adaptive Probabilistic Hough Transform \cite{yla-jaaski1994Adaptive} use a random selection of points from the point cloud instead of using all points, whereas the Randomized Hough Transform \cite{xu1990New} uses an approach where randomly, three points are selected from the point cloud.
These methods depend heavily on the choices of parameters, such as number of iterations or number of selected points. Another approach is based on clustering, such as the Depth Kernel-based Hough Transform (D-KHT). First, an attempt is made to cluster the point cloud into approximately co-planar clusters. This approach leads to an analytical solution with a reduced computational demand.

In this paper, we introduce how to combine the point cloud obtained from a LiDAR sensor and measurements from a microphone array in order to improve the performance of existing acoustic reflector detection methods. The planar surfaces that are estimated from the point cloud are used to reduce the influence of model mismatch, e.g., reflections from horizontal reflectors such as floors and ceilings. Moreover, we also propose a method to include the point cloud estimates as a priori information for solving the acoustic problem more robustly.

\section{Novel Acoustics-LiDAR reflector estimator}
We consider a smart loudspeaker system with a uniform circular array (UCA) of radius $r$ with $M$ microphones and a LiDAR sensor with a predefined Field of View (FOV) of $\beta_\text{hor}^\circ \times \beta_\text{ver}^\circ$. The coordinate system is defined such that the center of the microphone array, the loudspeaker point source, and the LiDAR sensor are at the origin. The angle $\alpha=0^\circ$ corresponds to the on-axis direction of the loudspeaker. The LiDAR is oriented such that the center of its FOV is at $180^\circ$. The method applies to any room type, but we use the shoe-box shaped room for ease of illustration.

First, the acoustic model and the inverse problem are stated. Then, a method to include horizontal reflectors by detecting horizontal planar surfaces from a LiDAR point cloud is given. Finally, the detected vertical planar surfaces from the point cloud are included as a priori information in the inverse problem.

\subsection{Acoustic Model and solution}
For the acoustic system to detect the walls, we assume that a discrete version of the received room impulse response from the loudspeaker to the microphone array is accurately known, denoted by $h[n,m]$, with discrete time-sample index $n\in\{0,\dots,N_h-1\}$ and microphone index $m\in\{0,\dots,M-1\}$, as well as the loudspeaker directivity and the microphone array geometry. Let $N_h$ denote the impulse response length. The measurements are sampled in time at $f_s$ Hz. The direct path of the loudspeaker is assumed to be removed from $h$ in a pre-processing stage. It is assumed that $h$ consists of first-order reflections due to the fact that only nearby reflectors are considered.

The Mirror Image Source Method (MISM) is used as a model to determine image source locations $s[q,m]$, using a radial sample index $q\in\{0,\dots,T-1\}$ on a grid in polar coordinates with radial step size $\Delta R=\frac{v_c}{f_s}$ and angular step size by $\Delta\alpha=\frac{2\pi}{M}$. The candidate locations are between $R_{\min}=R_a$ and $R_{\max}=\frac{Tv_c}{f_s}+R_a$, where $R_a$ denotes the UCA radius, $v_c$ the speed of sound and $T$ the integer number of radial grid points. $s[q,m]$ defines a partial impulse response, where the element $s[q,m]=1$ if an image source is present at index $[q,m]$. If an image source location is not present in the discrete set, it should be assigned to the closest grid point in the set.
 
\subsubsection{Inverse problem}
The problem can be posed as a linear system of equations \cite{zacca2021Inferring}. The candidate source positions for each of the $m$ angular directions, $\mathbf s^{(m)} = [s[0,m], \ldots, s[T-1,m]]^T$, are concatenated to express the  grid as
\begin{equation}\label{eq:s}
 \mathbf{s}=[[\mathbf{s}^{(0)}]^T,...,[\mathbf{s}^{(M-1)}]^T] ^T.
\end{equation}
The channel responses are arranged similarly, where $\mathbf h^{(m)} \in \mathbb R^{N_h}$:
\begin{equation}\label{eq:hvec}
 \mathbf{h}=[[\mathbf{h}^{(0)}]^T,...,[\mathbf{h}^{(M-1)}]^T] ^T.
\end{equation}
The model is then posed as:
\begin{equation}\label{eq:linearsystem}
 \mathbf{h}=\boldsymbol{\Phi}\mathbf{s}+\mathbf{n}
\end{equation}
where $\boldsymbol{\Phi}$ is defined by the known loudspeaker directivity and array configuration. Having defined this linear system from \autoref{eq:linearsystem}, it is possible to solve for $\mathbf{s}$. This is done by solving the minimization problem from \autoref{eq:optlinearsystem}, where $\lambda||\mathbf{s}||_1$ is introduced to promote sparsity. This inverse problem is computationally too expensive to extend to 3D using a spherical array. Furthermore, a spherical array is not feasible for practical loudspeaker design. Due to directivity, the loudspeaker driver emits maximum energy on-axis, making it challenging to detect reflectors in other directions in a noisy scenario, as well as in presence of other reflectors.

\begin{equation}\label{eq:optlinearsystem}
 \begin{aligned}
 \min_{\mathbf{s}} \quad & ||\boldsymbol{\Phi}\mathbf{s}-\mathbf{h}||^2_2+\lambda||\mathbf{s}||_1\\
 \end{aligned}
\end{equation}

\subsubsection{From Image Source Locations to Planar Surface Equation}
Once $\mathbf{s}$ is found, the next step is getting the planar surface equation parameters. 
From the peaks in $\mathbf s$, the corresponding distance $R_\text{MISM}$ and angle $\alpha_\text{MISM}$ are extracted.
The plane equation is described with the plane distance $\rho$ and the plane normal $\boldsymbol{\nu}=[\nu_x,\nu_y,\nu_z]$. Since the plane is located exactly halfway between the source and the image source, $\rho = \frac{R_\text{MISM}}{2}$, the plane normal values are $\boldsymbol{\nu}=[\cos\alpha_\text{MISM},\sin\alpha_\text{MISM},0]$.

\subsection{Pre-processing for detected horizontal reflectors from the point cloud}\label{subsec:horizontal}
The model mismatch from non-vertical structures complicates the recovery of walls. Given the location of a horizontal reflector, i.e., a floor or a ceiling, it can be accounted for in pre-processing by removing its response from $h[n,m]$. Using the LiDAR camera and the D-KHT algorithm\cite{vera2018hough}, the surfaces within the LiDAR FOV are found. If a plane where the $z$-component of the plane normal is non-zero is detected, its contribution is eliminated from the acoustic response $h[n,m]$.

There are two pieces of information known about these horizontal reflectors. The first one is that the time of arrival is equal at all microphones in the array. Secondly, the distance from the loudspeaker to the reflector is known from the LiDAR measurements. This distance is converted to the expected time delay for the impulse to arrive at the UCA.
The time delay cannot be used directly since delays from the loudspeaker response need to be considered. Here, this is done by taking the time difference from the peak of the direct path, which is known from the configuration. From this, the expected sample of the floor reflection is found. 

The goal is to exploit all this information to eliminate this reflection before employing the acoustic wall detection algorithm. We do this by minimizing the following optimization problem.
\begin{equation}\label{eq:flooropt}
    \begin{aligned}
        \min_{\mathbf{h}_{\text{hor}}} \quad & ||\mathbf{h}-\mathbf{I}_M\otimes \mathbf{h}_{\text{hor}}||^2_2\\
        \textrm{s.t.} \quad & ||\mathbf{L}\mathbf{h}_{\text{hor}}||_2^2\leq b\\
        \end{aligned}
\end{equation}
This optimization problem is constructed in such a way that both pieces of information are incorporated. $\mathbf{h}_{\text{hor}}$ is the response from the horizontal reflector that is estimated. Multiple reflectors can be estimated simultaneously. By using the Kronecker product with the identity matrix, it is ensured that the estimated horizontal response is the same for all microphone channels. The matrix $\mathbf L = \text{diag}(\mathbf l)$ is a weighting matrix. This weighting matrix is constructed such that the samples that lie far from the expected horizontal reflector location sample $p$, have a high weight, to limit the response of $\mathbf{h}_{\text{hor}}$ outside the area where the reflector is detected. The weighting factor scales with the squared distance to the expected location, i.e., $l[n] = |n-p|^2$. After minimizing the problem from \autoref{eq:flooropt}, the estimated horizontal response $\mathbf{h}_\text{hor}$ is subtracted from each response $\mathbf{h}^{(m)}$ as given in (\ref{eq:hmwalls}), where the remainder is the wall response that is required, i.e.,
\begin{equation}\label{eq:hmwalls}
    \mathbf{h}^{(m)}_{\text{walls}}=\mathbf{h}^{(m)}-\mathbf{h}_{\text{hor}}.
\end{equation}

\subsection{Including a priori information in the inverse problem}\label{subsubsec:lidarasaprior}
In addition to elimination of floor and ceiling reflections from $h$, we can also use the LiDAR to include additional a priori information on the positions of the image sources.
The LiDAR is positioned such that the direction where the loudspeaker emits less energy due to its directivity, is covered by its FOV. Now, first the plane detection from the point cloud using the D-KHT is performed. The output of this algorithm is a list of plane equations, consisting of the distance $\rho$ and the normal vector $\boldsymbol{\nu}$. From the normal $\boldsymbol{\nu}$, the angle in the \textit{xy}-plane at which the wall is located compared to the system is easily recovered $\alpha =\tan^{-1}(\frac{\nu_y}{\nu_x})$.
The acoustic algorithm solves for the image sources rather than the plane equation directly. The angle $\alpha_{\text{pc}}$ at which the plane from the point cloud is detected corresponds to the angle of the image source. The distance of an image source $\rho_{\text{MISM}}$ corresponds to twice the distance of the planar surface that corresponds to that image source, $R_{\text{MISM}} = 2\rho_{\text{pc}}$, where $\rho_{\text{pc}}$ is the distance from the system to the planar surface. Now the angle and the distance at which an image source is expected, if the planar surface is indeed a wall, are known. This is used as a prior in solving the inverse problem from \cite{zacca2021Inferring} in \autoref{eq:optlinearsystem}, by including it as a constraint. The optimization problem is then:

\begin{equation}\label{eq:directivity2opt}
    \begin{aligned}
        \min_{\mathbf{s}} \quad & ||\boldsymbol{\Phi}\mathbf{s}-\mathbf{h}_{\text{walls}}||^2_2+\lambda||\mathbf{s}||_1\\
        \textrm{s.t.} \quad & ||\mathbf{L}\mathbf{s}||_2^2\leq b.
    \end{aligned}
\end{equation}
Now, $\mathbf{L}$ is a diagonal matrix of size $MT\times MT$. This matrix is constructed by forming the vectors $\mathbf{l}^{(m)}$, that are constructed in a similar way as in \autoref{subsec:horizontal}. The vector $\mathbf{l}$ is the combination of all vectors $\mathbf{l}_{(m)}$, $[[\mathbf{l}^{(0)}]^T,...,[\mathbf{l}^{(M-1)}]^T]^T$. If the candidate location is the location that is found using the plane detection from the point cloud, its entry is zero. The entries are larger if the candidate location in the grid is further away from the expected location from the point cloud. More specifically, we set $l[m,n]$ to $l[m,n]=|n-n_{\text{pc}}|(1+|m-m_{\alpha_{\text{pc}}}|)$.

\section{Experimental Results}
\begin{figure}[H]
    \centering
    \begin{minipage}{0.45\linewidth}
    \centering
    \includegraphics[width=0.9\linewidth]{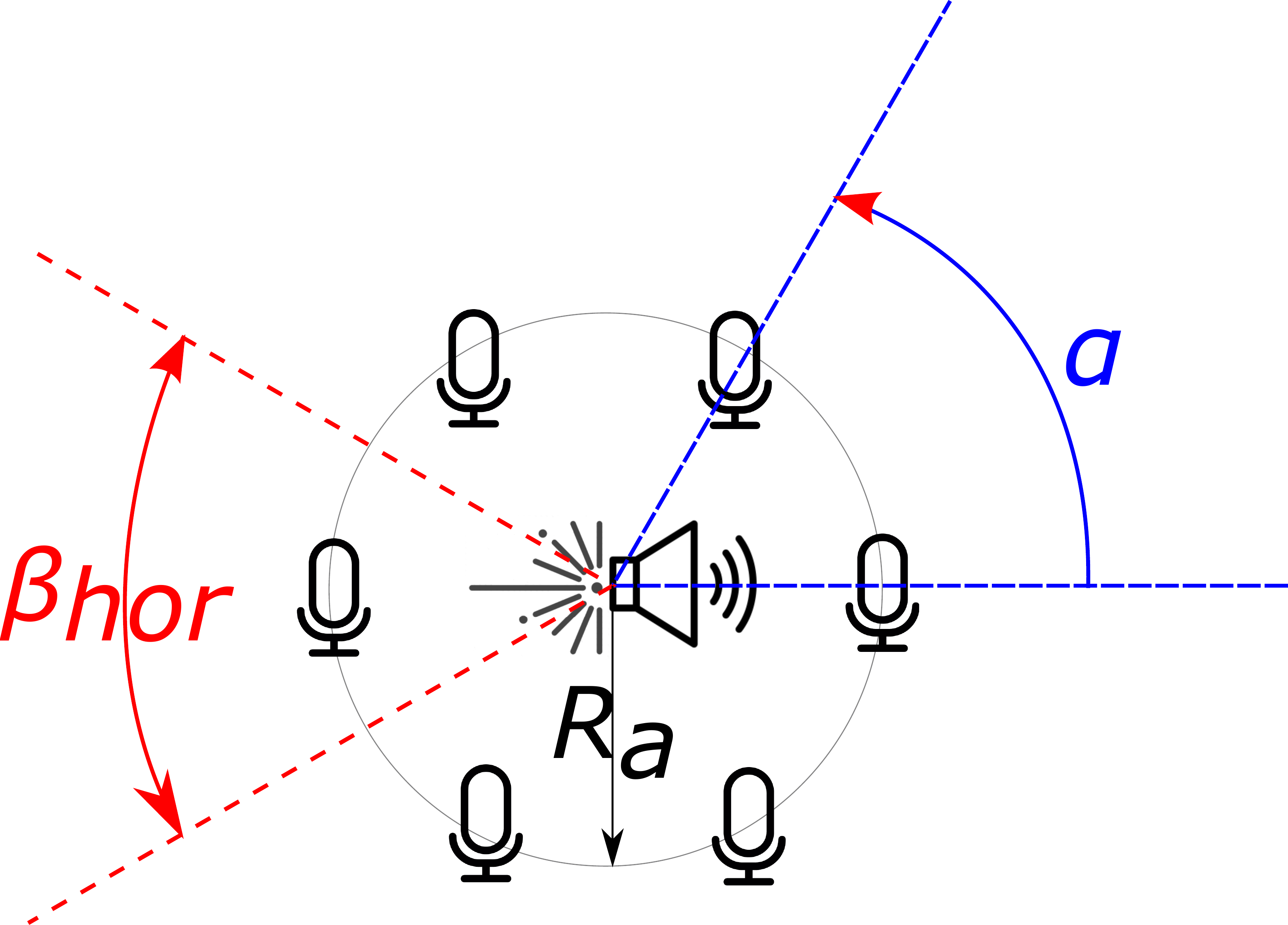}
    % \caption{Top view of system setup}
    \end{minipage}
    \begin{minipage}{0.45\linewidth}
    \centering
    \includegraphics[width=0.9\linewidth]{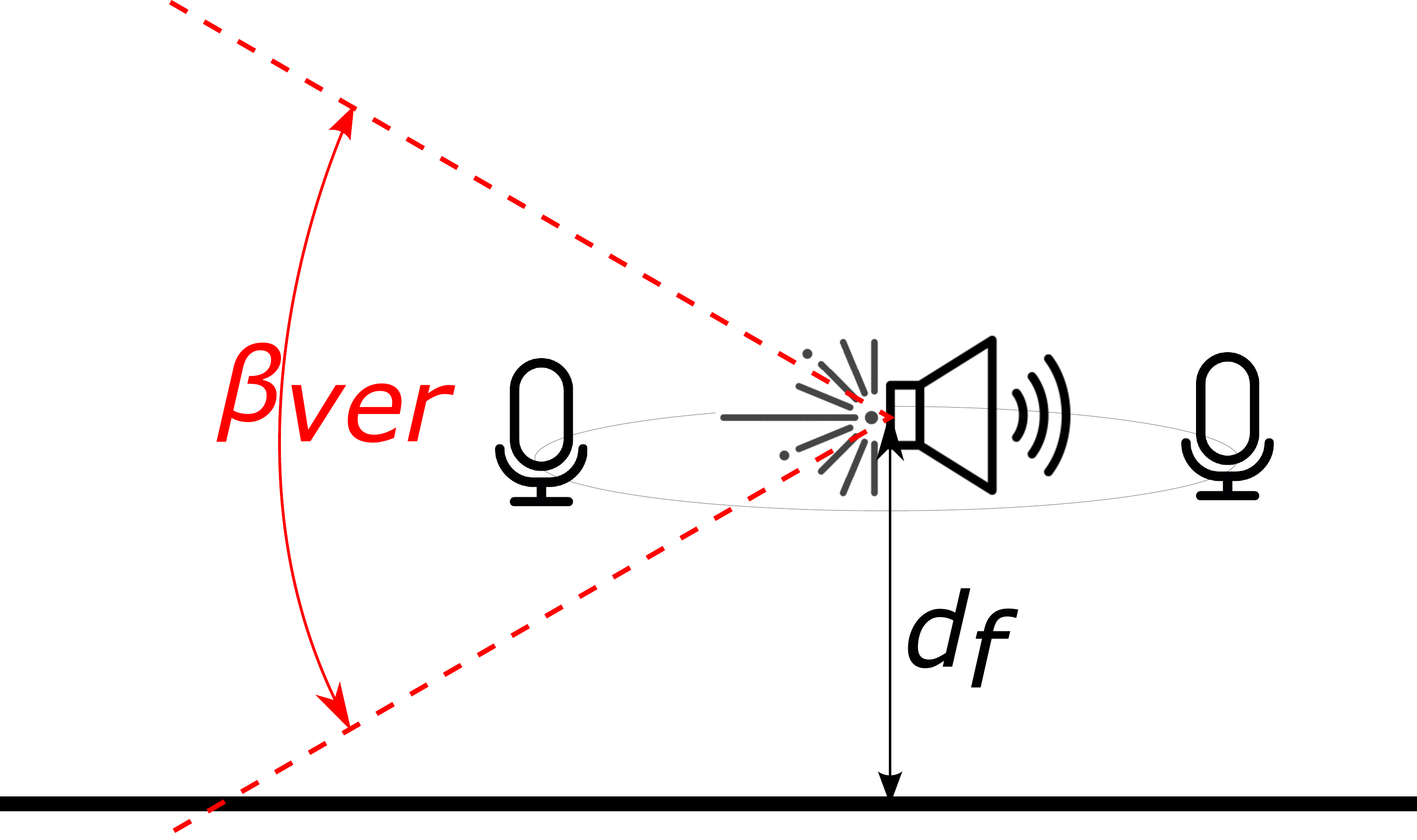}
    % \caption{Side view of system setup}
    \end{minipage}
    \caption{System setup. Left: Top view. Right: Side view.}
    \label{fig:systemsetup}
\end{figure}

In this section the proposed methods are evaluated in different scenarios.
The loudspeaker is modelled as a directive point source located at the origin where the front is positioned at $\alpha=0^\circ$, i.e., in the positive horizontal direction. The LiDAR sensor is also located in the origin, but its front is directed towards the negative horizontal direction $\alpha=180^\circ$. We denote its FOV by $\beta_{\text{hor}}\times \beta_{\text{ver}}$. A UCA containing $M$ microphones surrounds the loudspeaker and the LiDAR sensor. The set-up of the co-located system is shown in \autoref{fig:systemsetup}.

\subsection{Experiment 1 — Detection of a wall in presence of a floor}\label{sec:resultexp3}
\begin{figure}[H]
    \centering
    \includegraphics[width=0.6\linewidth]{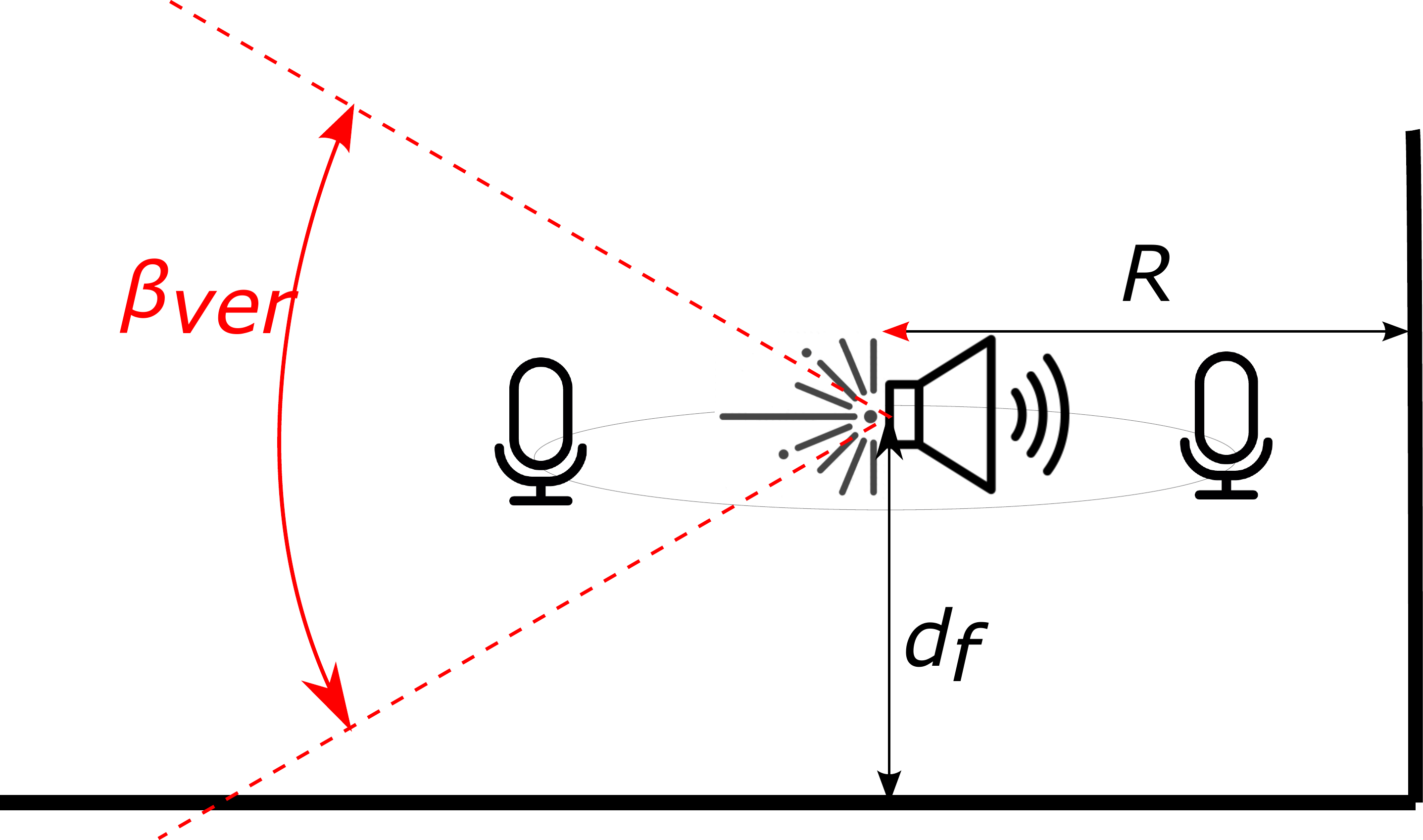}
    \caption{Side view of the set-up for Experiment 1. The co-located system is placed at height $d_{\text{f}}$ at a distance $R=1$ m from a wall at angle $\alpha=0^\circ$.}
    \label{fig:setupexp3}
\end{figure}

A significant ambiguity is introduced when reflections that were not included in the model are present. The purpose of this experiment is to show that the proposed method reduces this ambiguity. The co-located system is placed in front of a wall at an angle of $\alpha = 0^\circ$ and at the distance $R=1$ m. A floor is introduced at varying distance $d_{\text{f}}$, as shown in \autoref{fig:setupexp3}. The loudspeaker is assumed to be omnidirectional and the LiDAR FOV is now $50^\circ\times70^\circ$. This scenario is noiseless.

\begin{figure}
    \centering
    \includegraphics[trim=70 0 30 0, clip, width=\linewidth]{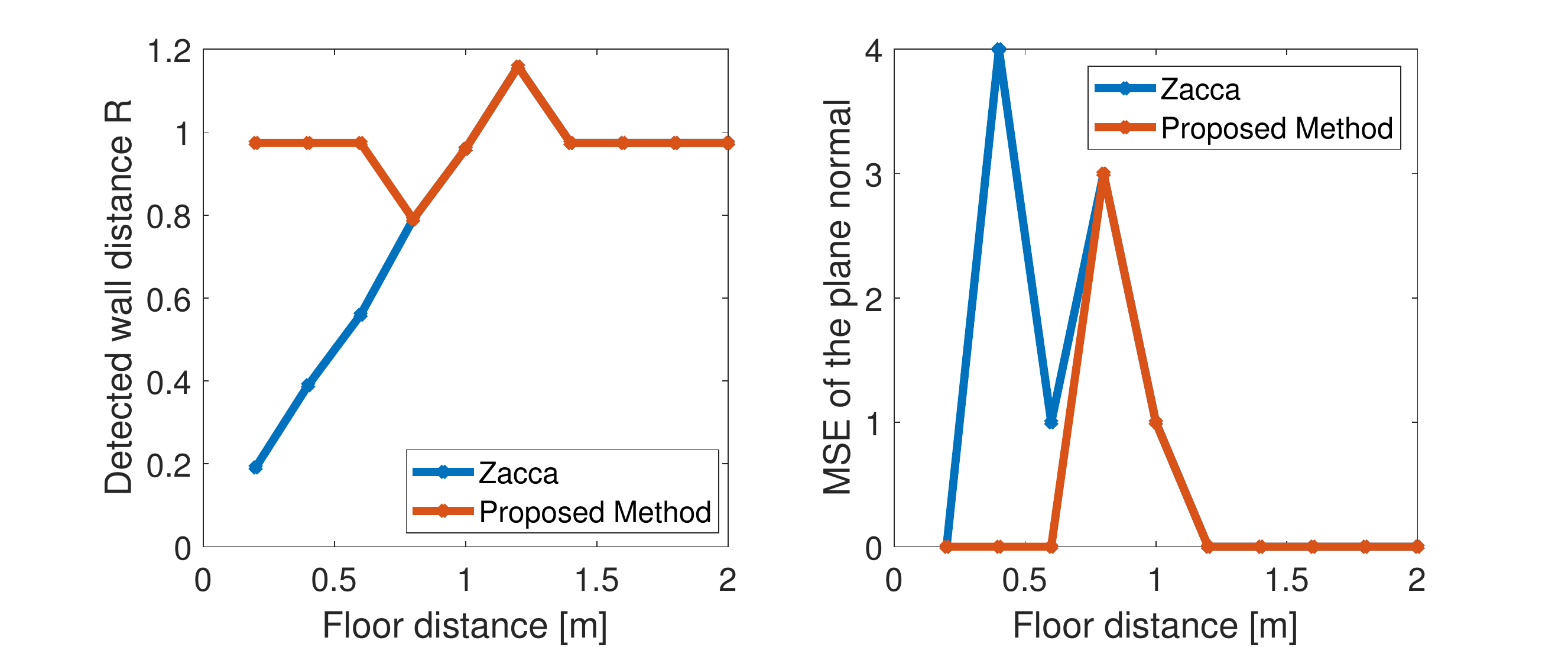}
    \caption{Left: The detected distance $R$ for the method of Zacc\'a \cite{zacca2021Inferring} and the proposed method at different floor heights. The ground truth is $R=1$. Right: The mean square error with the ground truth normal vector for Zacc\'a's method and the proposed method at different floor heights.}
    \label{fig:resexp3}
\end{figure}

In \autoref{fig:resexp3}, the results of this experiment are shown. In the first figure, the distance that is detected using either Zacc\'{a}'s algorithm \cite{zacca2021Inferring} or the proposed method is given. In the second figure, the mean square error (MSE) of the plane normal is given. 
From this simulation, what is seen is that using Zacc\'a's algorithm, the first arriving reflection is detected. When a floor is present closer than the wall, a wall is detected at this distance, where the angle is unpredictable. When the floor is further away than the wall, the wall can be found reliably. 
Using the proposed method, it is possible to eliminate this effect. What is seen is that when the floor is within the LiDAR sensor FOV, the detection can be done reliably. Above a distance of $0.6$ m, the floor gets below the LiDAR sensor's FOV and the same results as with Zacc\'a's method are achieved. The additional steps in the processing increases the computational complexity. However, the steps are less demanding than if the linear system would be extended to 3D and the practical implications of a LiDAR sensor are smaller than those of a spherical array.

\subsection{Experiment 2 — Single wall scenario}\label{sec:resultexp1}
\begin{figure}
    \centering
    \includegraphics[width=0.5\linewidth]{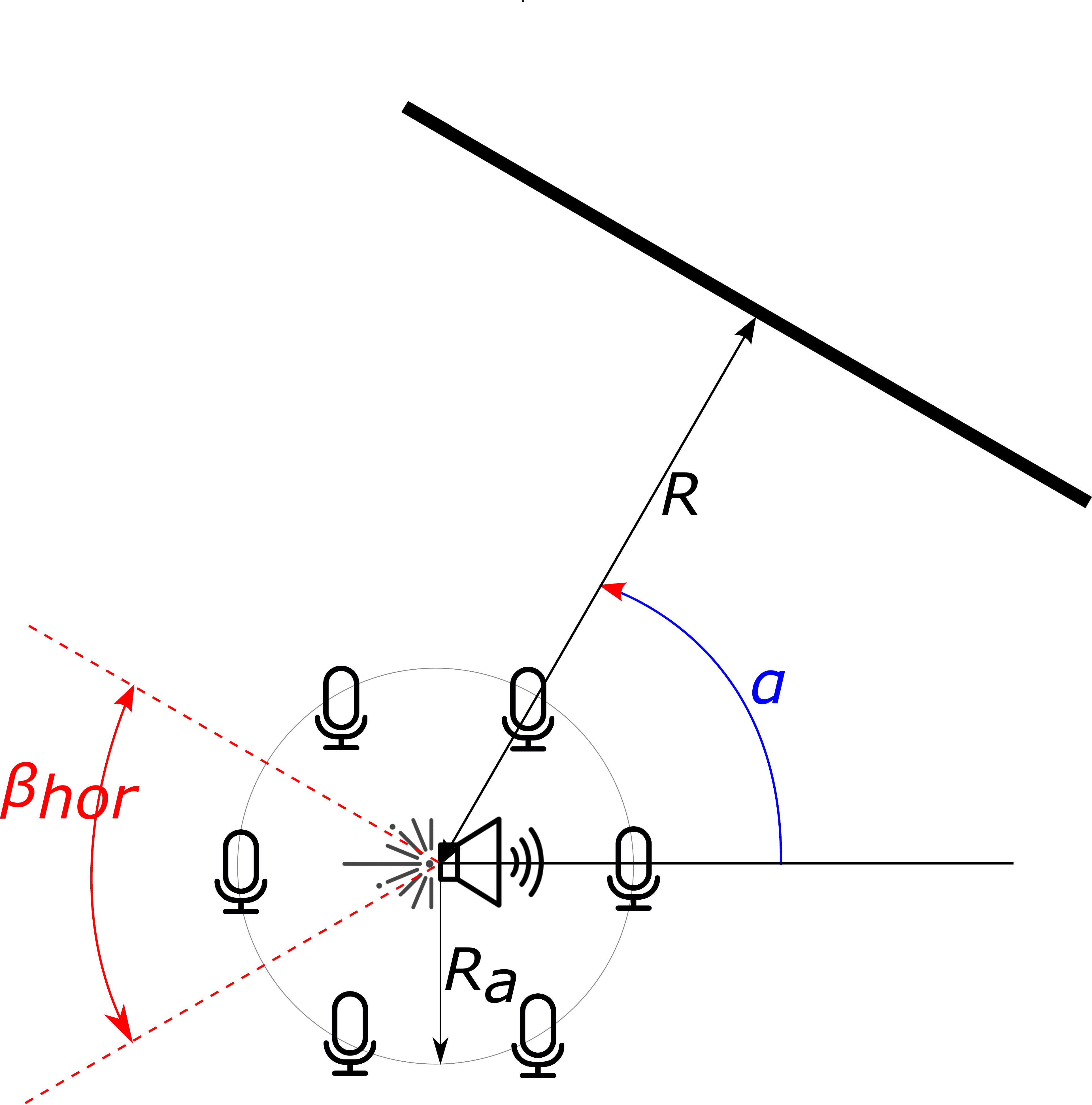}
    \caption{The setup used for experiment 2. The thick black line illustrates a wall that is placed at different angles $\alpha$ around the system.}
    \label{fig:setupexp1}
\end{figure}

The purpose of this experiment is to evaluate the performance of the proposed method compared to the state-of-the-art method \cite{zacca2021Inferring}, where specifically angles of $90^\circ < \alpha <270^\circ$ are of interest.
This is demonstrated with a single wall scenario. This wall is placed at 0.5 m and is rotated around the co-located system, as shown in \autoref{fig:setupexp1}. 

\begin{figure}
    \centering
    \includegraphics[width=0.5\linewidth]{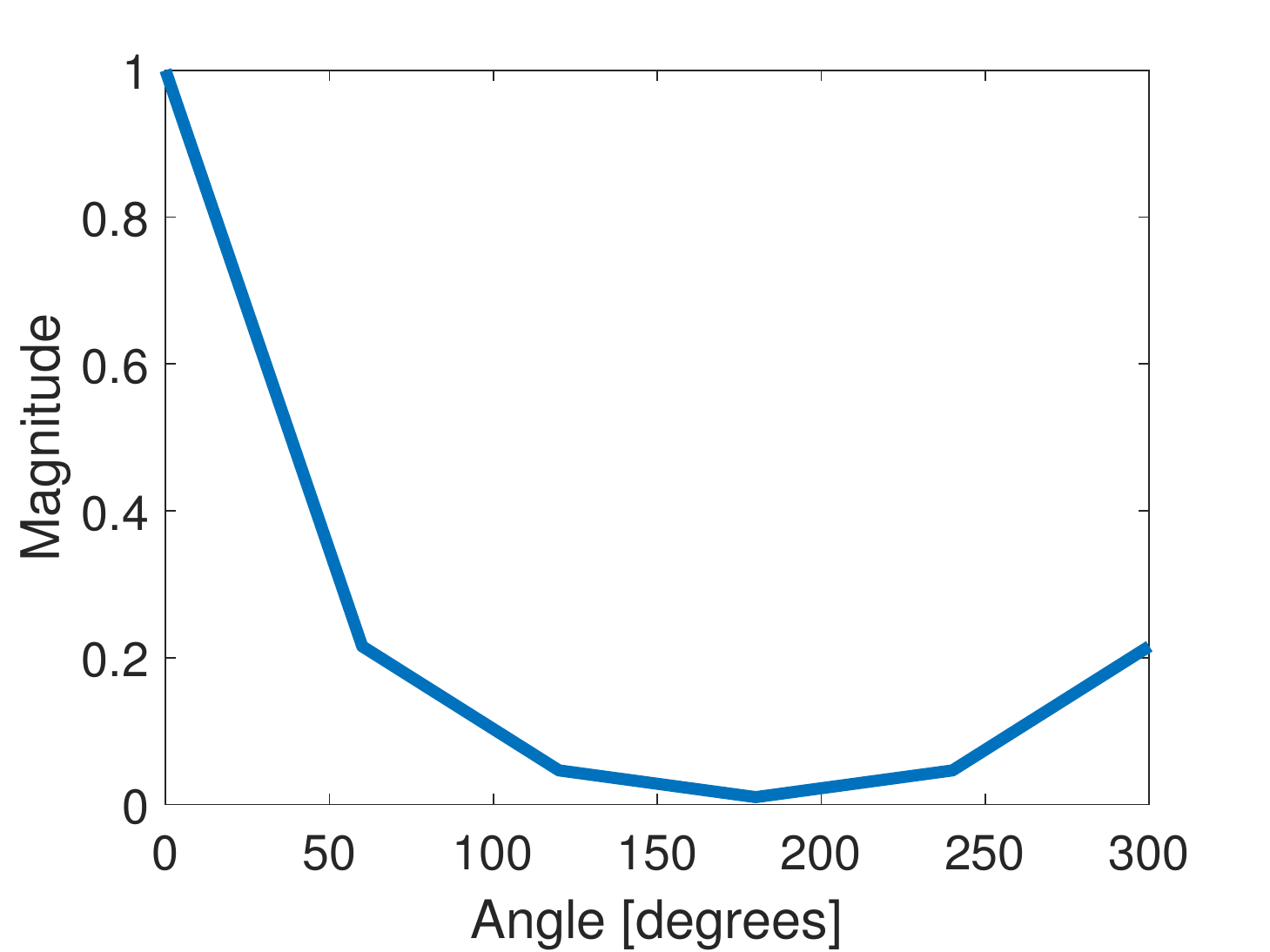}
    \caption{Directivity characteristic of the loudspeaker.}
    \label{fig:magnitudedirectivity}
\end{figure}

The field of view of the LiDAR is $70^\circ\times 50^\circ$ and is centered at $180^\circ$. The loudspeaker directivity characteristic is given in \autoref{fig:magnitudedirectivity}. For simplicity, only magnitude scaling is used.

\begin{figure}
    \centering
    \begin{minipage}{0.45\linewidth}
    \centering
    \includegraphics[trim=25 0 30 0, clip, width=\textwidth]{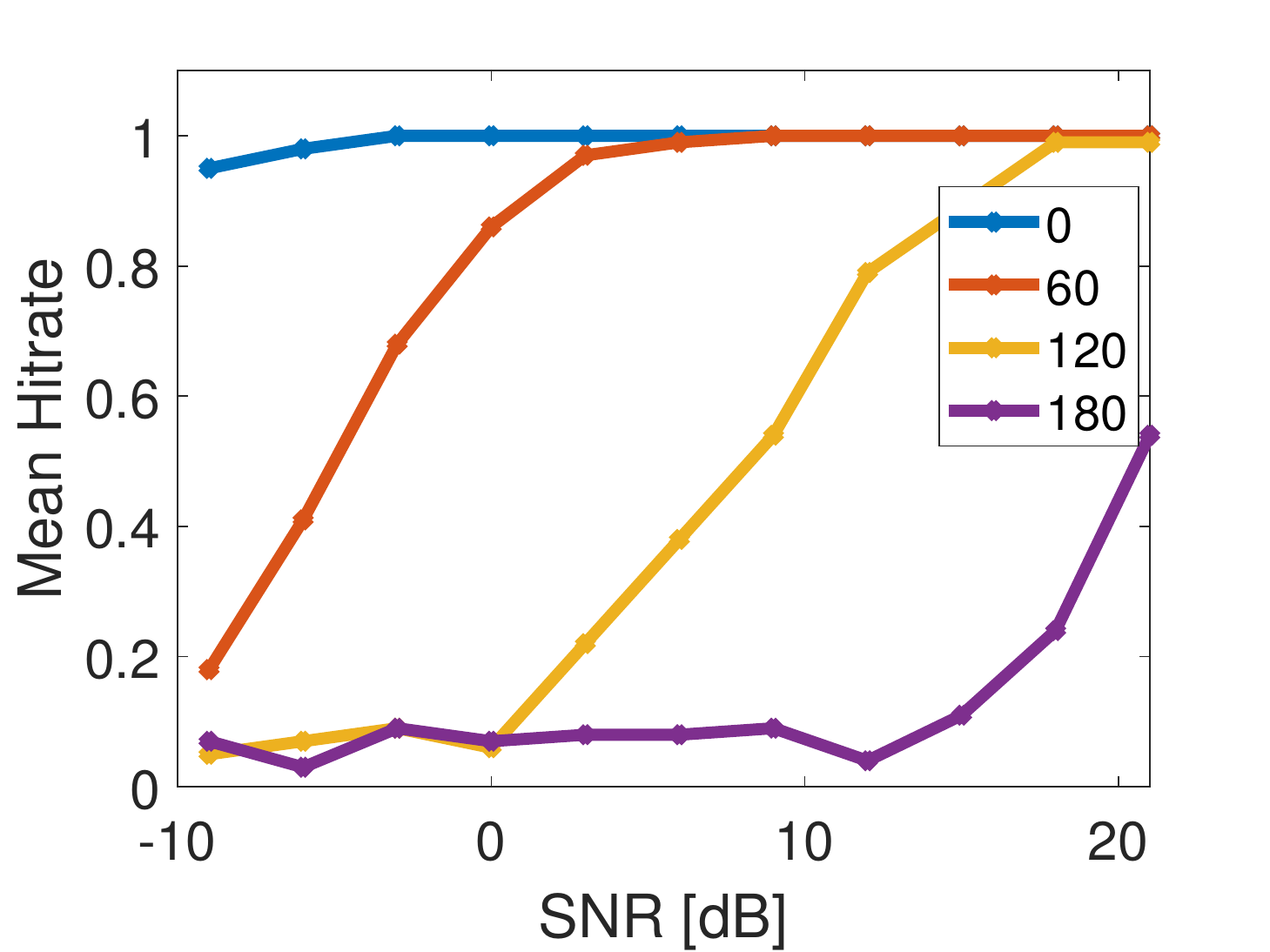}
    \end{minipage}
    \begin{minipage}{0.45\linewidth}
    \centering
    \includegraphics[trim=25 0 30 0, clip, width=\textwidth]{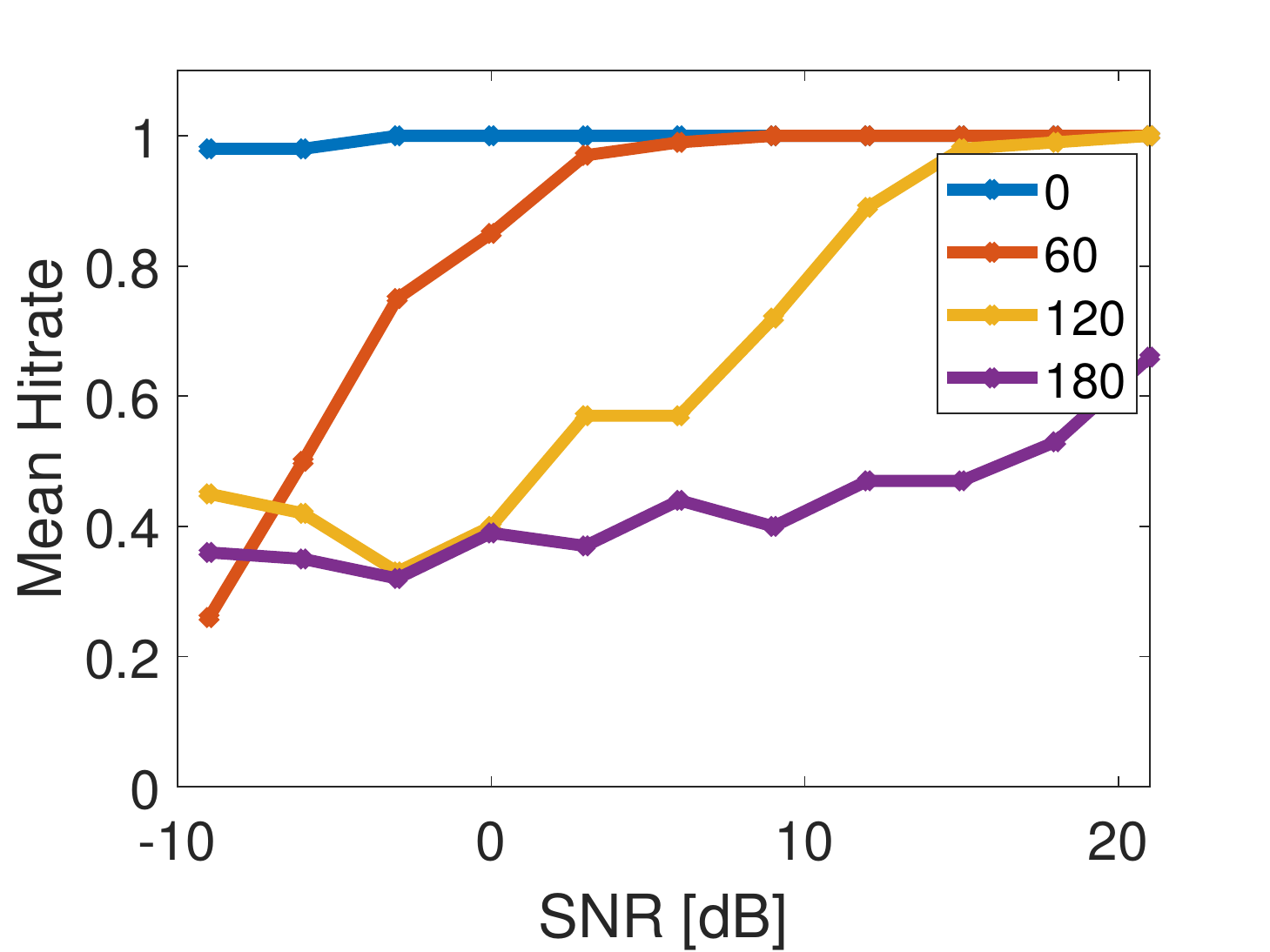}
    \end{minipage}
    \caption{Detection of a rotating wall around the co-located system. Left: Zacc\'{a}'s method \cite{zacca2021Inferring}. Right: Proposed method}
    \label{fig:resexp1}
\end{figure}

The LiDAR measurements are noiseless and the acoustic SNR is varied from -9 dB to 21 dB. A Monte-Carlo simulation of 100 runs is performed. 
In \autoref{fig:resexp1}a, the mean hitrate of Zacc\'{a}'s method is given for each angle and for different signal-to-noise ratio (SNR) values. The hitrate is 1 if a wall is detected correctly and averaged over the runs. 
What can be seen from this figure, is that it is challenging to detect a wall at a low SNR, and that it becomes difficult as the angle increases to $180^\circ$.
In \autoref{fig:resexp1}b, the mean hitrate, when the proposed method is used, is given for each angle and for different SNR values. 
Here, a significant improvement is achieved for the angles in the LiDAR sensor FOV, i.e., $120^\circ$ and $180^\circ$. 

\subsection{Experiment 3 — Scenario with windows}\label{sec:resultexp2}
\begin{figure}[]
    \centering
    \includegraphics[width=0.6\linewidth]{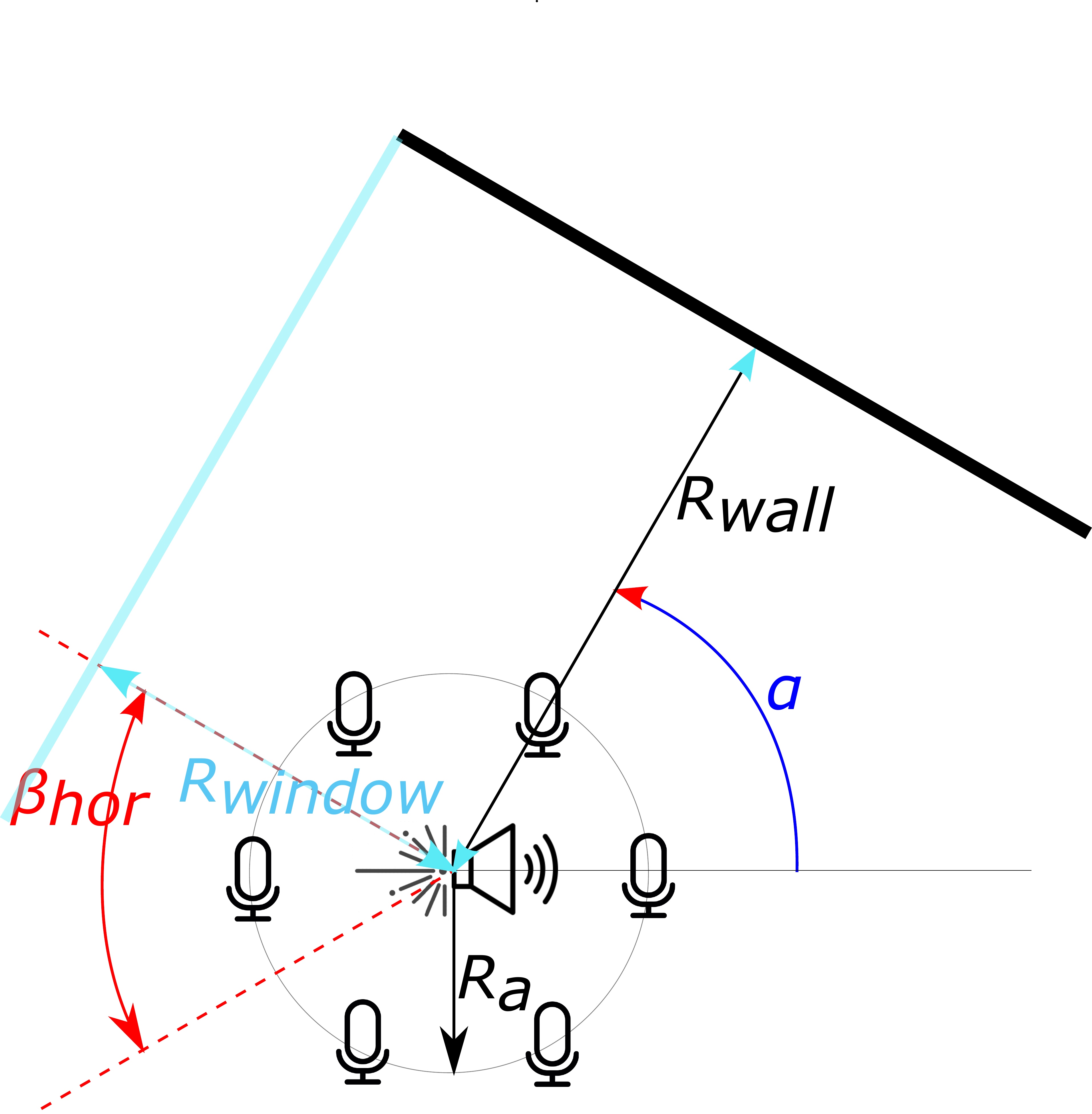}
    \caption{The setup used for experiment 2. The thick black line illustrates a wall that is placed at different angles $\alpha$ around the system. The blue line represents a window at angle $90^\circ$ with the wall.}
    \label{fig:setupexp2}
\end{figure}

Using the LiDAR sensor, it is challenging to detect windows. In this experiment, the consequence is shown in a dual wall scenario. One of the reflectors is a window, i.e., its response is present in the RIR, but not in the point cloud. Again, the configuration is rotated around the co-located system as demonstrated in \autoref{fig:setupexp2}.

The distance $R_\text{window}=0.6$ m and the distance $R_\text{wall}=0.4$ m. The FOV of the LiDAR is again $70^\circ\times50^\circ$ and is centered at $180^\circ$. The loudspeaker directivity from \autoref{fig:magnitudedirectivity} is used. 

\begin{figure}[]
    \centering
    \includegraphics[trim=65 0 0 0, clip, width=\linewidth]{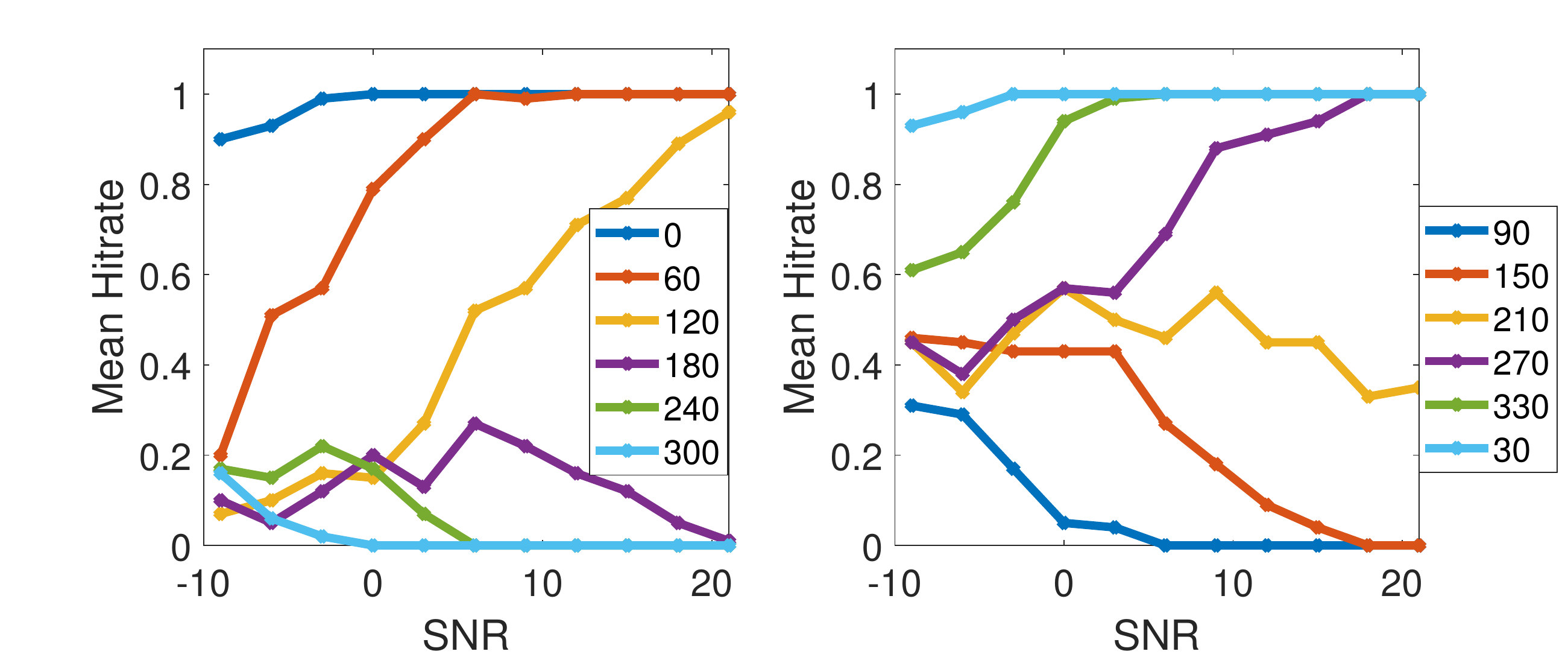}
    \caption{Detection of a wall and a window at different angles using Zacca's method. Left: Wall detection accuracy. Right Window detection accuracy.}
    \label{fig:resexp2acoustics}
\end{figure}

The LiDAR measurements are noiseless and the acoustic SNR is varied from -9 dB to 20 dB. A Monte-Carlo simulation of 100 runs is performed. 
In \autoref{fig:resexp2acoustics}, the mean hitrate of Zacc\'{a}'s method is given for each angle and for different SNR values. For a hitrate of 1, the wall or window is detected correctly. 
The detectability of the wall remains similar to the scenario in \autoref{fig:resexp1}a, for angles $0^\circ <\alpha<120\circ$. Here, the window is placed at an angle with a lower magnitude response than the wall. Once the window is at an angle with a larger magnitude response, the wall recovery degrades. Due to the simultaneous detection, it can be more challenging to detect the surface in the direction with a lower magnitude response, since the detection relies on finding the maxima in $\mathbf{s}$. Often, two such maxima are detected next to each other for one image source.

\begin{figure}[]
    \centering
    \includegraphics[trim=70 0 0 0, clip, width=\linewidth]{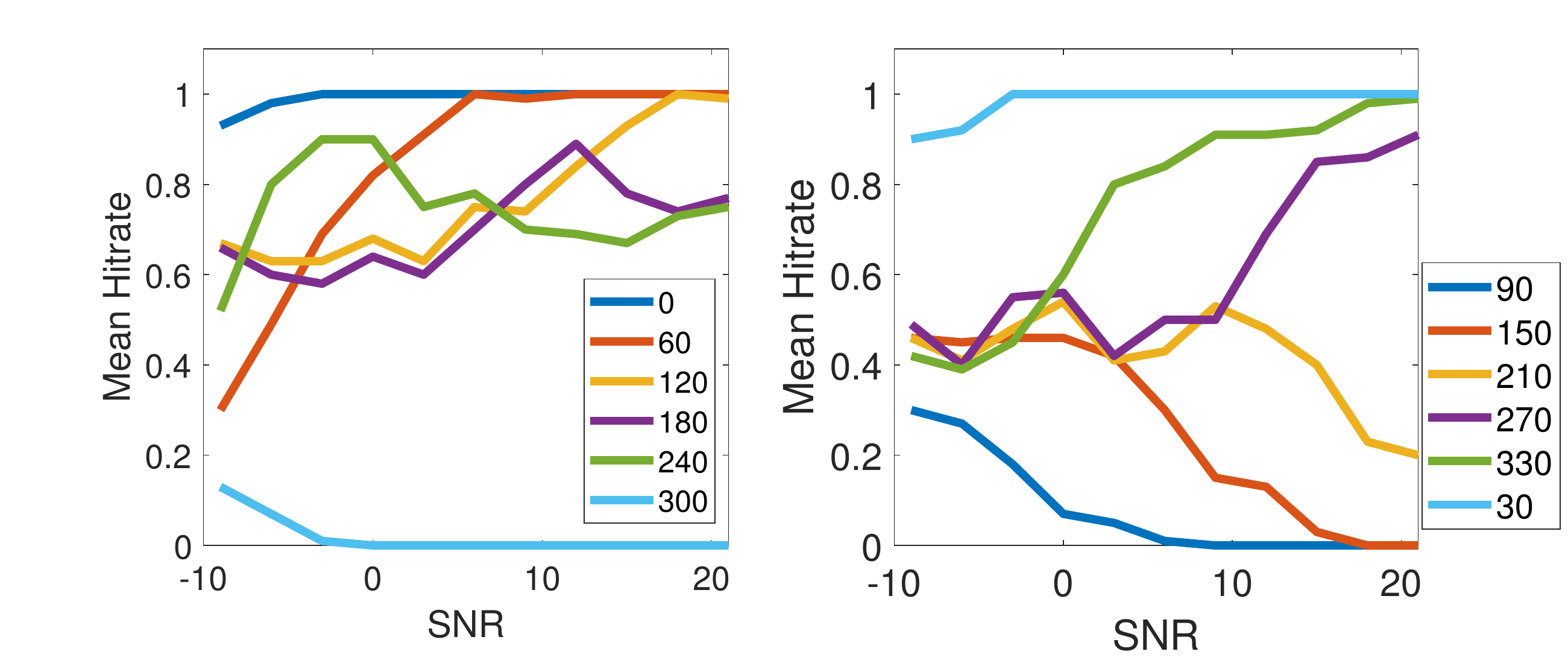}
    \caption{Detection of a wall and a window at different angles using the proposed method. Left: Wall detection accuracy. Right Window detection accuracy.}
    \label{fig:resexp2method2}
\end{figure}

Now, the experiment is repeated with the proposed method. The results are presented in \autoref{fig:resexp2method2}. The detection of the wall improves at angles $120-240^\circ$ with the proposed method, while the detection of the window remains similar. At wall angle $240^\circ$, the hitrate improves, whereas the corresponding window at $330^\circ$ is less detectable. Again, this is due to the simultaneous estimation; two maxima are detected next to each other in $\mathbf{s}$ for one image source.

\section{Conclusions}
The presented method exploits the information of the additional LiDAR sensor to improve the robustness and accuracy of existing acoustic reflector detection methods. First, using an omnidirectional model, it is demonstrated how detecting a horizontal reflector from the point cloud enables us to eliminate its negative effect. 
Then, the detection of vertical planar surfaces from the point cloud is included in the acoustic inverse problem as a priori information. It was shown that combining the two sensing modalities leads to a better performance of surface detection in low-energy directions.

\bibliographystyle{IEEEtran}
\bibliography{mscthesis.bib}

\end{document}